\documentclass[11pt]{article}
\usepackage[utf8]{inputenc}
\usepackage[english]{babel}
\usepackage[normalem]{ulem}
\usepackage{amsmath, amssymb, amsfonts, amsthm}
\usepackage{multirow}
\usepackage{booktabs}
\usepackage[text={16.3cm,22.5cm}]{geometry}
\usepackage{graphicx}
\usepackage[dvipsnames]{xcolor}
\usepackage{slashed}
\usepackage{mathrsfs}
\usepackage{yfonts}
\usepackage{hyperref}
\usepackage{cite}
\usepackage{verbatim}
\hypersetup{colorlinks,bookmarksopen,bookmarksnumbered,citecolor=blue,
	linkcolor=black,pdfstartview=FitH,urlcolor=blue}

\newcommand{\Tr}[0]{\mathop{\mathrm{Tr}}}

\begin{document}
	
	\title{\vspace{-0.8cm}
		{\normalsize
			\flushright TUM-HEP 1313/20\\}
		\vspace{1cm}
		\bf   Neutrino parameters in the Planck-scale lepton
		number breaking scenario with extended scalar sectors
		\\ [8mm]}
	
	\author{Cesar Bonilla$^1$, Johannes Herms$^{2,3}$, Alejandro Ibarra$^2$, Patrick Strobl$^2$,\\[2mm]
		{\normalsize\it $^1$ Departamento de Física, Universidad Católica del Norte,}\\[-0.05cm]
		{\normalsize\it  Avenida Angamos 0610, Casilla 1280, Antofagasta, Chile}\\[2mm]
		{\normalsize\it $^2$ Physik-Department, Technische Universit\"at M\"unchen, }\\[-0.05cm]
		{\normalsize\it James-Franck-Stra\ss{}e, 85748 Garching, Germany}
		\\[2mm]
		{\normalsize\it $^3$ Max-Planck-Institut f\"ur Kernphysik, }\\[-0.05cm]
		{\normalsize\it Saupfercheckweg 1, 69117 Heidelberg, Germany}
	}

	\date{}
	\maketitle
	\thispagestyle{empty}
	\vskip 1.5cm
	\begin{abstract}
		Two-loop effects on the right-handed neutrino masses can have an impact on the low-energy phenomenology, especially when the right-handed neutrino mass spectrum is very hierarchical at the cut-off scale. In this case, the physical masses of the lighter right-handed neutrinos can be dominated by quantum effects induced by the heavier ones. Further, if the heaviest right-handed neutrino mass is at around the Planck scale,  two-loop effects on the right-handed neutrino masses generate, through the seesaw mechanism, an active neutrino mass which is in the ballpark of the experimental values. In this paper we investigate extensions of the Planck-scale lepton number breaking scenario by additional Higgs doublets (inert or not). We find that under reasonable assumptions these models lead simultaneously to an overall neutrino mass scale and to a neutrino mass hierarchy in qualitative agreement with observations. 
	\end{abstract}
	
	\newpage

\section{Introduction}
\label{sec:intro}

The Standard Model (SM) predicts that neutrinos are massless particles. However, neutrino oscillation experiments have established that at least two active neutrinos have a tiny, albeit non-zero, mass (for a recent determination of the neutrino parameters from a global fit to oscillation experiments, see \cite{deSalas:2020pgw}).
The simplest scenario that explains the tininess of the neutrino masses is the  so-called type-I seesaw mechanism~\cite{Minkowski:1977sc,Yanagida:1979as,Mohapatra:1979ia,Schechter:1980gr}. In this scenario, the SM particle content is extended by at least two spin-1/2 particles, singlet under the Standard Model gauge group. The gauge symmetry allows a Yukawa coupling of the fermion singlets to the Standard Model Higgs and lepton doublets, $Y$, which leads after the breaking of the electroweak symmetry to a Dirac neutrino mass; for this reason, the fermion singlets are also known as right-handed neutrinos (RHNs). The gauge symmetry also allows a Majorana mass for the RHNs, $M$, which breaks lepton number by two units. This mass is unrelated to the electroweak symmetry breaking scale, $v=246$ GeV, and  can take \textit{a priori} any value between 0 and the cut-off scale of the model, usually taken to be the Planck scale. The seesaw mechanism assumes $M\gg Y v$, leading in turn to active neutrino masses  $m_\nu\sim  Y^2 v^2/M$, which are suppressed with respect to the electroweak symmetry breaking scale by the small factor $v/M$, thus explaining the smallness of neutrino masses.

The seesaw mechanism provides a qualitative explanation for the smallness of the neutrino masses, but not quantitative, since the Yukawa couplings and the RHN masses are free parameters of the model. It has been conjectured that some RHN masses could be at the Planck scale, from the fact that Planck-scale physics is expected to break all global quantum numbers~\cite{Akhmedov:1992hh,Berezinsky:2004zb}. Regardless of possible theoretical motivations, it is worthwhile from the phenomenological standpoint to entertain this possibility, since the number of free parameters of the model is then somewhat reduced. 

In a tree-level analysis, the Planck-scale RHNs do not seem to contribute significantly to the low-energy neutrino phenomenology, since $v^2/M_{\rm P}\sim 10^{-5}$ eV. Nevertheless, it was pointed out in Ref.~\cite{Ibarra:2018dib,Ibarra:2020eia} that quantum effects can dramatically alter the conclusions when the RHN masses are very hierarchical at the cut-off scale. Since the total lepton number is broken already, there is no symmetry protecting the lighter right-handed neutrino masses against quantum effects. The latter can then receive sizable  contributions from two-loop quantum effects (possibly dominant), thereby affecting  the low-energy phenomenology. Interestingly, when the lepton number is broken at around the Planck scale, this scenario predicts an overall neutrino mass scale in the ballpark of the experimental values. We denominate this scenario as Planck-scale lepton number breaking scenario. The predicted neutrino mass hierarchy, however, is typically much larger than the measured value, as generically expected in the type-I seesaw mechanism~\cite{Casas:2006hf}, unless the parameters take special values (not necessarily fine-tuned). The observed mild neutrino mass hierarchy could then be regarded as a hint that the type-I seesaw model must be extended, and in particular its realization with Planck-scale lepton number breaking. 

A simple extension of the seesaw model consists in adding to the particle content a scalar particle with identical gauge quantum numbers as the Standard Model Higgs doublet. The extra doublet, even if it is heavy, can play an important role in the low-energy neutrino phenomenology. Concretely, if the active neutrino mass hierarchy is very large at the decoupling scale of the lightest right-handed neutrino, quantum effects on the active neutrino masses induced by the second Higgs doublet can generate a mild neutrino mass hierarchy, in agreement with observations~\cite{Ibarra:2011gn,Grimus:1999wm}. 

In a variant of this model, so-called ``scotogenic", the fermion singlets and the extra scalar doublets are furnished with a $Z_2$ symmetry. Then, neutrino masses are not generated at tree level, but at the one-loop level~\cite{Ma:2006km}. Further, the model contains one dark matter candidate, usually the $Z_2$-odd scalar (the ``inert" doublet), or in some regions of the parameter space the singlet fermion. The predicted mass hierarchy among the active neutrinos is again too large in general. However, a mild mass hierarchy can be generated in the presence of more than one $Z_2$-odd  inert doublet~\cite{Hehn:2012kz,Escribano:2020iqq}.

In this paper we will analyze the Planck-scale lepton number breaking scenario, which is successful in predicting the correct overall neutrino mass, with an extended scalar sector, which is successful in explaining the observed mild neutrino mass hierarchy. We will show that, under plausible assumptions, it is possible to reproduce simultaneously the correct neutrino mass scale and mass hierarchy. In Section~\ref{sec:RGEs} we calculate the quantum effects on the right-handed neutrino parameters in the two-Higgs doublet model, and  in Section~\ref{sec:RHmasses} we determine the mass spectrum of right-handed neutrinos in the Planck-scale lepton number breaking scenario. In Section~\ref{sec:LHmasses} we explore the implications for the active neutrino masses in this scenario, and in Section~\ref{sec:scotogenic} for its scotogenic variant. Finally, in Section~\ref{sec:conclusion} we present our conclusions.

\section{Quantum effects on the right-handed neutrino mass matrix in the  two-Higgs doublet model}
\label{sec:RGEs}
We consider in this section the two-Higgs doublet model (2HDM) extended by three RHNs, $N_i$, $i=1,2,3$.
The scalar potential reads
\begin{align}
	\label{eq:V_2HDM}
	V =
	\sum_{a,b} \mu_{ab}^2 \Phi_a^\dagger \Phi_b
	+ \sum_{a,b,c,d} \lambda_{abcd} \left( \Phi_a^\dagger \Phi_b \right)
	\left( \Phi_c^\dagger \Phi_d \right)\;,
\end{align}
where $\Phi_a$, $a=1,2$, are scalar $SU(2)$-doublets with hypercharge $Y=1$.
The hermiticity of the potential requires $\mu_{ab}^2=(\mu_{ba}^2)^*$ and $\lambda_{abcd}=\lambda_{badc}^*$ (for a comprehensive review of the 2HDM, see \cite{Branco:2011iw}).
The part of the Lagrangian involving the RHNs reads:
\begin{align}
 \mathcal{L}_N = \frac{1}{2} \overline{N_i} i \slashed{\partial} N_i - \frac{1}{2} M_{ij} \overline{N^c_i}N_j 
 -Y^{(a)}_{\alpha i}\overline{L_\alpha} N_i \widetilde\Phi_{a} + \mathrm{h.c.}, \label{eq:lagrangianN}
\end{align} 
where $L_\alpha$ ($\alpha = e, \mu, \tau$) are the lepton doublets. Further, 
$\widetilde\Phi_{a} = i\sigma_2\Phi_a^{*}$ denotes the charge conjugated scalar fields, and $N_i^c=-i\gamma^2 N_i^*$ the charge conjugated RHN fields.

The parameters of the Lagrangian in Eq.~(\ref{eq:lagrangianN})  are subject to quantum corrections, which can have significant impact on the phenomenology.
The leading quantum effects can be calculated using the renormalization group equations (RGEs). Including up to two-loop effects, the RGE of the RHN mass matrix reads:
\begin{align}
\label{eq:RGEforM}
\frac{d M}{d \log\mu} =\sum_{a,b} \left(M Q^{(ab)} + Q^{(ab) T} M +4 P^{(ba) T} M P^{(ab)}\right),
\end{align}
where for convenience, we have defined
\begin{align}
P^{(ab)} &= \frac{1}{16 \pi^2} Y^{(a)\dagger} Y^{(b)},\\
\label{eq:quantities}
Q^{(ab)} &= \left(1+\mathcal{G}\right)  P^{(ab)} \delta_{ab} - \frac{1}{4} P^{(ab)} P^{(ba)}- \left(\frac{9}{2} \frac{\mathrm{Tr}\left[Y_u^{(a)}Y_u^{(b)\dagger}\right]}{16 \pi^2} + \frac{3}{2} \textrm{Tr}\left[ P^{(ba)}\right]\right) P^{(ab)},\\
\mathcal{G} &= \frac{1}{16\pi^2}\left( \frac{17}{8} g_1^2 + \frac{51}{8} g_2^2 \right).
\end{align}
Here, $g_1$ and $g_2$ are the $U(1)_Y$ and $SU(2)_L$ gauge couplings and $Y_u^{(a)}$ are the up-quark Yukawa coupling matrices to both Higgs doublets (we assume that the Yukawa couplings of the other SM fermions to both Higgs doublets are negligible).
For the purposes of this paper, it is sufficient to consider the one-loop RGE of the neutrino Yukawa coupling. Using \texttt{SARAH}~\cite{Staub:2008uz}, we obtain:
\begin{align}
	(16\pi^2)\frac{dY^{(a)}}{d\log\mu}=& \left[3 \sum_{b}\textrm{Tr}\left(Y_u^{(a)} Y_u^{{(b)}\dagger}\right) +\sum_{b}\textrm{Tr}\left(Y^{(a)}Y^{(b)\dagger}\right)-\frac{3}{4} g_1^2 -\frac{9}{4} g_2^2 \right]Y^{(a)}\nonumber \\
	&+\sum_{b}Y^{(b)}Y^{(b)\dagger}Y^{(a)}+ \sum_{b}\frac{1}{2} Y^{(a)}Y^{(b)\dagger}Y^{(b)}\;.
\end{align}
We will work in the basis where the RHN mass matrix is real and diagonal at the cut-off energy scale $\Lambda$:
\begin{align}
\label{eq:massmatrixR}
 M(\mu)\Big|_{\mu=\Lambda} = \begin{pmatrix} M_1 & 0 & 0 \\ 0 & M_2 & 0 \\ 0 & 0 & M_3 \end{pmatrix}\;.
\end{align}
Integrating Eq.~(\ref{eq:RGEforM}), one can calculate  the RHN mass matrix at the scale $\mu < \Lambda$. Keeping  terms up to the order $\mathcal{O}\big(P^{(ab)2}\big)$ we obtain:
\begin{align}
\label{eq:Mlowscale2ndorder}
M(\mu) &\simeq \left[ 1 +  \sum_{a} \left(P^{(aa)} t + \frac{1}{2} P^{(aa) 2}t^2\right) 
  \right]^T M\left[ 1 +  \sum_{b}\left(  P^{(bb)}  t + \frac{1}{2} P^{(bb)2} t^2\right)    \right] \\ \nonumber
&+ 4 \sum_{ab}  P^{{(ba)}T} M  P^{(ab)} t  
+ \mathcal{O}\left(P^{(ab)3}\right)\;,
\end{align}
where we have denoted $t = \mathrm{log}(\mu/\Lambda)$.

\begin{figure}[t]
	\begin{center}
		\includegraphics[width=0.45\textwidth]{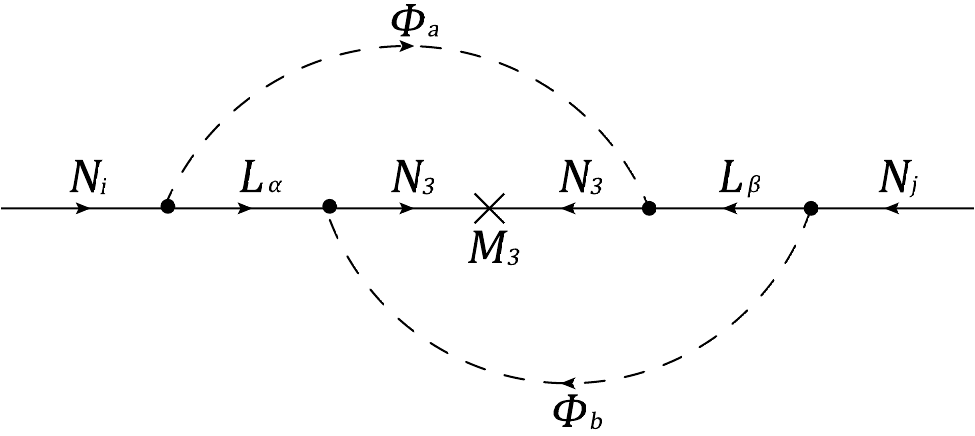}
	\end{center}
	\caption{Leading two-loop diagram generating radiatively right-handed neutrino masses from the breaking of lepton number by $M_3$.}
	\label{fig:2Ldiag}
\end{figure}

We are interested in the scenario where the mass matrix at the cut-off scale is approximately rank-1, $M_1, M_2 \ll M_3$ and where $M_3\sim  M_{\rm P}$, being $M_{\rm P}=1.2\times 10^{19}$ GeV the Planck mass.  To emphasize the main features of the RGE, let us consider the limiting scenario where $M_1=M_2=0$, namely when the mass matrix is exactly rank-1 (our conclusions, however, apply to a wider class of scenarios, as we will discuss below). One can readily check that at $\mathcal{O}\left(P^{(ab)}\right)$, {\it i.e.} keeping just the first  line of Eq.~(\ref{eq:Mlowscale2ndorder}), the mass matrix at the scale $\mu$ is also rank-1. However, at $\mathcal{O}\left(P^{(ab)2}\right)$, the mass matrix in general becomes  rank-3: the RGE evolution generates radiatively non-zero values for $M_{1,2}(\mu)$ proportional to $M_3$, through the diagram shown in Fig.~\ref{fig:2Ldiag}. This effect was explored in~\cite{Ibarra:2020eia} for the seesaw scenario with one Higgs doublet (see also ~\cite{Aparici:2012vx}). In that case, however, a rank-1 mass matrix at the scale $\Lambda$ remained rank-1 at order $\mathcal{O}(P)$, became rank-2 at order $\mathcal{O}(P^2)$, and became rank-3 only at order $\mathcal{O}(P^4)$. However, the existence of an additional RHN Yukawa coupling in the 2HDM (and thereby the existence of additional flavor symmetry breaking parameters), allows to increase the rank of the mass matrix at lower order in perturbation theory. Here, we have considered the limiting case where the mass matrix is exactly rank-1 at the cut-off scale. For an approximately rank-1 mass matrix, one finds that the physical masses of the two lightest RHNs can be dominated by the quantum contribution induced by the heaviest RHN. Correspondingly, their tree-level masses would not play any role in the phenomenology, thus rendering a more predictive scenario.

In order to construct the low-energy effective theory of the Planck-scale lepton number breaking scenario, we first integrate out the heaviest RHN at the scale $\mu=M_3$. The effective Lagrangian reads:
\begin{align}
\label{eq:Leff}
\mathcal{L}_{\mathrm{eff}} \simeq\frac{1}{2} \frac{Y_{\alpha 3}^{(a)} Y_{\beta 3}^{(b)}}{M_{33}}\Big|_{\mu=M_3} \left(\overline{L_\alpha} \widetilde\Phi_a\right)\left(\widetilde{\Phi}_b^{T} L_\beta^c \right) - \mathbb{Y}^{(a)}_{ \alpha i} \overline{L_\alpha } \widetilde{\Phi}_a N_i -\frac{1}{2} \mathbb{M}_{ij} \overline{N_i^c} N_j + \mathrm{h.c.},
\end{align}
with Yukawa and mass matrices given by 
\begin{align}
\mathbb{Y}_{\alpha i}^{(a)} &\simeq \left(Y_{\alpha i}^{(a)} - \frac{M_{i 3} Y_{\alpha 3}^{(a)}}{M_{33}}\right)\Big|_{\mu=M_3},\nonumber\\
\mathbb{M}_{ij} &\simeq \left(M_{ij} - \frac{M_{i 3} M_{j 3}}{M_{33}}\right)\Big|_{\mu=M_3},
\label{eq:eff_parameters}
\end{align}
with $i,j = 1,2$. 

The first term in Eq.~(\ref{eq:Leff}) is a Weinberg operator giving rise to a contribution to the active neutrino masses suppressed by $M_3$. In the Planck-scale lepton number breaking scenario $M_3$ is close to the Planck scale. Hence, this term gives a negligible contribution to the neutrino  masses. We will drop this term henceforth, and describe the effective theory as a two-right handed neutrino model with the Lagrangian:
\begin{align}
	\label{eq:Leff2}
	\mathcal{L}_{\mathrm{eff}} \simeq - \mathbb{Y}^{(a)}_{ \alpha i} \overline{L_\alpha} \widetilde{\Phi}_a N_i -\frac{1}{2} \mathbb{M}_{ij} \overline{N_i^c} N_j + \mathrm{h.c.}\;,
\end{align}
with $\mathbb{Y}$ and $\mathbb{M}$ given in Eq.~(\ref{eq:eff_parameters}). Using Eq.~(\ref{eq:Mlowscale2ndorder}), and using that $M_{i3}/M_{33}|_{\mu=M_3}$ is a small parameter, one can further approximate
\begin{align}
	&\mathbb{Y}_{\alpha i}^{(a)}\simeq Y_{\alpha i}^{(a)}\;, \nonumber \\
	&\mathbb{M}_{ij} \simeq -4 M_3 \,\mathrm{log}\left(\frac{\Lambda}{M_3}\right) \sum_{a,b} P_{3i}^{(ba)} P_{3j}^{(ab)}\;,
	\label{eq:eff_parameters_approx}
\end{align}
where we have assumed generic Yukawa couplings at the scale $\Lambda$, and that the running does not significantly modify the Yukawa couplings. However, two-loop quantum effects lift the zeroes in the RHN mass matrix and generate radiatively two mass eigenvalues. The implications for the mass spectrum of right-handed neutrinos and active neutrinos will be discussed in the next sections.

\section{Right-handed neutrino masses}
\label{sec:RHmasses}

Let us first calculate the mass spectrum of heavy neutrinos in our scenario.
The eigenvalues can be calculated from Eq.~(\ref{eq:eff_parameters_approx}) using the tensor invariants\footnote{In this paper we assume all parameters to be real for simplicity. In the complex case, the corresponding invariants are  $I_1 = \mathrm{Tr}\big[\mathbb{M}^\dagger \mathbb{M}\big] = M_1^2+M_2^2$ and $I_2 = \mathrm{det}\big[\mathbb{M}^\dagger \mathbb{M}\big] = M_1^2\,M_2^2$.} 
\begin{align}
\label{eq:invariants}
I_1 &= \mathrm{Tr}\big[\mathbb{M}\big] =M_1+M_2\Big|_{\mu=M_3} ,\nonumber\\
I_2 &= \mathrm{det}\big[\mathbb{M}\big] =M_1 M_2 \Big|_{\mu=M_3} \;.
\end{align}
Assuming a hierarchy between the eigenvalues, one obtains
\begin{align}
\label{eq:RHmassinvariants}
M_2 \Big|_{\mu = M_3} &\simeq I_1,\nonumber \\ 
M_1 \Big|_{\mu= M_3} &\simeq \frac{I_2}{I_1}\;,
\end{align}
which are complicated expressions that depend on the Yukawa couplings.

In order to gain some analytical understanding of the results, let us consider for simplicity rank-1 Yukawa matrices. The Yukawas $Y^{(a)}$ can be expressed in terms of the non-vanishing eigenvalues, $y_a$, and the tensor products of two vectors in flavor space $\vec{u}^{(a)}_L$, $\vec{u}_R^{(a)}$ (normalized to unity):
\begin{align}
\label{eq:rank1yuk}
 Y^{(a)} =  y_a \vec{u}^{(a)}_L \otimes \vec{u}_R^{(a)T}.
\end{align}
In a similar notation, the RHN neutrino mass matrix can be cast as:
\begin{align}
	\label{eq:rank1mass}
	M(\Lambda) = M_3 \, \vec{\omega} \otimes \vec{\omega}^T\;.
\end{align}

Clearly, the physical RHN masses can only depend on invariant quantities. In our simplified scenario, we have three vectors in the RHN flavor space, $\vec \omega$ and $\vec u_R^{(a)}$, and two vectors in the left-handed neutrino (LHN) flavor space, $\vec u_L^{(a)}$. There are then only four invariant quantities related to the relative orientation among these vectors, $(\vec u_R^{(1)}\cdot \vec \omega)$, $(\vec u_R^{(2)}\cdot \vec \omega)$,  $(\vec u_R^{(1)}\cdot \vec u_R^{(2)})$ and $(\vec u_L^{(1)}\cdot \vec u_L^{(2)})$, as well as the three eigenvalues $y_1$, $y_2$ and $M_3$. In terms of these invariants, the radiatively generated RHN masses $M_1$ and $M_2$ at the scale $\mu=M_3$ read:

\begin{align}
M_2\Big\vert_{\mu = M_3} &\simeq  -\frac{4 M_3 \mathrm{log}\big(\frac{\Lambda}{M_3}\big)}{\big(16 \pi^2\big)^2}
\sum_{a,b} y_a^2 y_b^2\, \big(\vec u_R^{(a)}\cdot \vec\omega\big)\,
\big(\vec u_R^{(b)}\cdot \vec\omega\big)\,
\big(\vec u_L^{(a)} \cdot \vec u_L^{(b)} \big)^2
\Big[1-\big(\vec u_R^{(a)}\cdot \vec\omega\big)\big(\vec u_R^{(b)}\cdot \vec\omega\big)\Big]\;,\nonumber \\
M_1\Big\vert_{\mu = M_3} &\simeq -\frac{4 M_3 \mathrm{log}\big(\frac{\Lambda}{M_3}\big)}{\big(16 \pi^2\big)^2}  \frac{ y_1^4  y_2^4 \, (\vec u_R^{(1)}\cdot \vec\omega)^2\,
	(\vec u_R^{(2)}\cdot \vec\omega)^2\,\left( \vec \omega \cdot \big(\vec u_R^{(1)} \times \vec u_R^{(2)}  \big) \right)^2 \big[1-
	\big(\vec u_L^{(1)} \cdot \vec u_L^{(2)}\big)^4\big]
}{\sum_{a,b} y_a^2 y_b^2\, (\vec u_R^{(a)}\cdot \vec\omega)\,
	\big(\vec u_R^{(b)}\cdot \vec\omega\big)\,
	\big(\vec u_L^{(a)} \cdot \vec u_L^{(b)} \big)^2
	\Big[1-\big(\vec u_R^{(a)}\cdot \vec\omega\big)\big(\vec u_R^{(b)}\cdot \vec\omega\big)\Big]} \;,
\label{eq:RHmass}
\end{align}
where the square of the triple product explicitly reads:
\begin{align}
 \left(\vec \omega \cdot \big(\vec u_R^{(1)} \times \vec u_R^{(2)}  \big)\right)^2&=1-\big(\vec u^{(1)}_R\cdot\vec\omega\big)^2-\big(\vec u^{(2)}_R\cdot\vec\omega\big)^2-\big(\vec u^{(1)}_R\cdot u^{(2)}_R\big)^2+2\big(\vec u^{(1)}_R\cdot\vec\omega\big)\big(\vec u^{(2)}_R\cdot\vec\omega\big)\big(\vec u^{(1)}_R\cdot u^{(2)}_R\big)\;.
\end{align}

From these equations one concludes that in order to generate a non-vanishing $M_2$ either $\vec u_{R}^{(1)}$ or $\vec u_{R}^{(2)}$ must be non-orthogonal to $\vec \omega$. Further, in order to generate a non-vanishing $M_1$, the three following conditions must be  simultaneously fulfilled: {\it i}) both  $\vec u_{R}^{(1)}$ and $\vec u_{R}^{(2)}$ must be non-orthogonal to $\vec \omega$, {\it ii}) $\vec u_{R}^{(1)}$ and $\vec u_{R}^{(2)}$ must point in different directions and  {\it iii}) $\vec u_{L}^{(1)}$ and $\vec u_{L}^{(2)}$ must also point in different directions. In more generality, generating $M_2$ requires at least two independent directions in the RHN flavor space, and generating $M_1$ requires three independent directions in the RHN flavor space, as well as two independent directions in the LHN flavor space. This can also be understood from the breaking of the global flavor group, $U(3)_L\times U(3)_R\rightarrow$ nothing, by the Yukawa couplings~\cite{Ibarra:2018dib}. In the Standard Model extended with RHNs, a rank-1 Yukawa matrix and a rank-1  mass matrix provide two directions in the RHN flavor space, and therefore this model generates only $M_2$, but not $M_1$ (due to a residual global $U(1)$ symmetry). A rank-2 Yukawa matrix can generate via quantum effects a non-vanishing $M_1$, although suppressed by the next-to-largest Yukawa eigenvalue and only beyond two-loop order. In the 2HDM extended with RHNs there are many more directions in flavor space, and therefore it is possible to generate radiatively both $M_1$ and $M_2$ with rank-1 Yukawa couplings. From these expressions, one can also construct the limit where one of the RHNs, say $N_1$, has no coupling to the left-handed doublets; this would correspond to a model with only two RHNs. In this case, $\vec u_R^{(1)}$, $\vec u_R^{(2)}$ and $\vec\omega$ are coplanar and therefore $M_1|_{\mu=M_3}=0$.

For our analysis we will find convenient to use as invariants the following four angles:
\begin{align}
	\vec u_L^{(1)} \cdot \vec u_L^{(2)} &= \cos\theta_L, \nonumber \\ 
	\vec u_R^{(1)} \cdot \vec u_R^{(2)} &= \cos\theta_R, \nonumber \\ 
	\vec u_R^{(1,2)} \cdot \vec \omega &= \cos\theta_{1,2},
	\label{eq:angleDefinitions}
\end{align}
as well as the three eigenvalues $y_1$, $y_2$ and $M_3$. With this parametrization, the eigenvalues in Eq.~(\ref{eq:RHmass}) can be written as
\begin{align}
	M_2\bigg\vert_{\mu = M_3} &\simeq - \frac{4  y_1^2 y_2^2}{\left(16 \pi^2\right)^2}M_3 \mathrm{log}\left(\frac{\Lambda}{M_3}\right) \left[\left(\frac{y_1}{y_2} s_1 c_1 \right)^2 +\left(\frac{y_2}{y_1}  s_2 c_2\right)^2+ 2  c_L^2 c_1c_2 \left(c_R-c_1c_2\right)\right], \nonumber \\
	M_1\bigg\vert_{\mu = M_3} &\simeq-\frac{4  y_1^2 y_2^2}{\left(16 \pi^2\right)^2}M_3 \mathrm{log}\left(\frac{\Lambda}{M_3}\right)
	 \frac{c_1^2 c_2^2\big( 1 - c_L^4 \big) \left(1-c_1^2 - c_2^2 - c_R^2 + 2 c_1 c_2 c_R\right)  }{\left(\frac{y_1}{y_2} s_1 c_1 \right)^2 +\left(\frac{y_2}{y_1}  s_2 c_2\right)^2+ 2  c_L^2 c_1c_2 \left(c_R-c_1c_2\right)},
	 \label{eq:RHmass_at_M3}
\end{align}
with $s_i = \sin \theta_i$ and $c_i = \cos\theta_i$ ($i = 1,2,L,R$). 

Below the scale $M_3$, both RHN masses are subject to additional quantum effects, although in this case they amount to small corrections. Therefore, one can approximate the physical masses for $N_1$ and $N_2$ by the running masses at the scale $\mu=M_3$ in Eq.~(\ref{eq:RHmass_at_M3}).

\begin{figure}[t!]
	\begin{center}
		\includegraphics[width=0.6\textwidth]{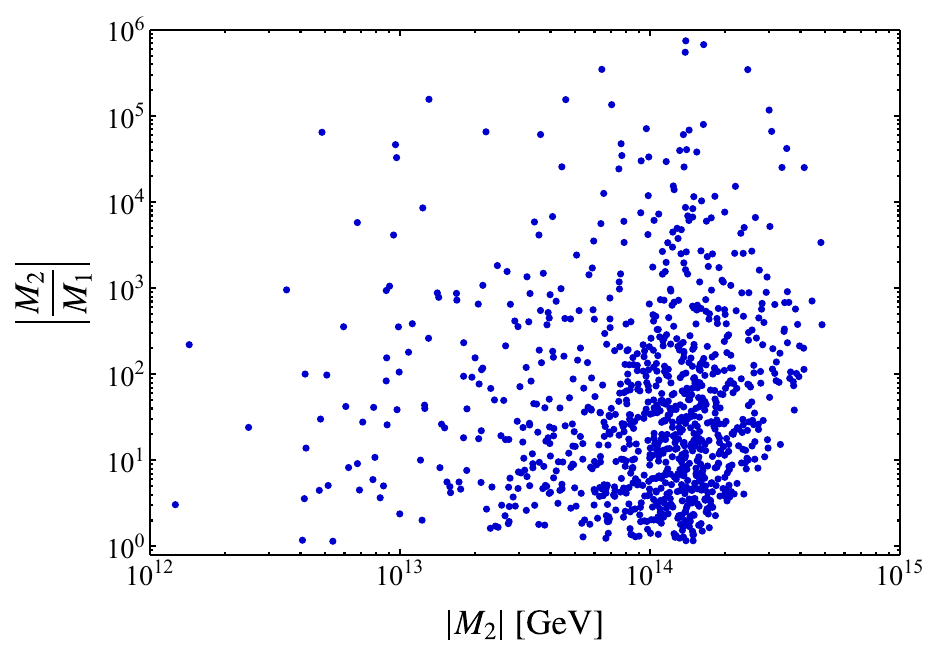}
	\end{center}
	\caption{Scan plot showing the mass hierarchy $\vert M_2/M_1 \vert$ vs. $|M_2|$ for the radiatively generated masses in the Planck-scale lepton number breaking scenario with two Higgs doublets,  assuming $\Lambda=M_{\rm P}$,  $M_3 = M_{\rm P}/\sqrt{8 \pi}$, $y_1=y_2=1$ and random angles between 0 and $2 \pi$.}
	\label{fig:RHN_hierarchy}
\end{figure}

The overall mass scale of both $M_1$ and $M_2$ is determined by the parameter
\begin{align}
	M_0\equiv \frac{4 y^2_1  y^2_2 }{\big(16 \pi^2\big)^2}    M_3 \mathrm{log}\left(\frac{\Lambda}{M_3}\right)\;.
	\label{eq:M0}
\end{align}
Numerically, in the Planck-scale lepton number breaking scenario
\begin{align}
M_0 \sim 2 \times 10^{15}\,{\rm GeV} \, y^2_1 y^2_2 \, \left(\frac{M_3}{M_{\rm P}}\right) \,\log\left(\frac{M_3}{M_{\rm P}}\right)\;,
\end{align}
which is generically at the seesaw scale. The concrete values of $M_1$ and $M_2$ depend on complicated combinations of angles and $y_1/y_2$.  We show in Fig.~\ref{fig:RHN_hierarchy} a scan plot with the ratios of the physical masses $M_2/M_1$ vs. $M_2$ from solving the RGEs numerically between the cut-off scale $\Lambda$ and the scale $\mu=M_3$. For the plot, we have taken $\Lambda=M_{\rm P}$, $M_3=M_{\rm P}/\sqrt{8\pi}$, $M_1=M_2=0$, $y_1=y_2=1$ and random angles $\theta_L,\theta_R,\theta_{1,2}$ at the cut-off. One concludes from the Figure that for $y_1\sim y_2\sim 1$, two-loop quantum effects generate non-vanishing values for $M_1$ and $M_2$, with $M_2\sim 10^{14}$ GeV and $M_2/M_1$ typically smaller than $\sim 100$. 

Let us note that the same conclusion holds whenever the physical masses of $N_1$ and $N_2$ are dominated by quantum contributions proportional to $M_3$, even if they do not vanish at the cut-off scale. In this case, quantum effects can milden the hierarchy between the two lighter eigenvalues, leading to $M_2|_{\mu=M_3}\sim M_1|_{\mu=M_3}$. The consequences for the light neutrino mass spectrum are expected to be dramatic. If two-loop effects had been neglected and $M_1 \ll M_2$ at the decoupling scale, the generation of a mild mass hierarchy $m_3\sim m_2$ would look rather accidental. However, the quantum effects induced at two loops by the two Higgs doublets generically lead to a mild hierarchy between $M_1$ and $M_2$, and therefore it will be easier to generate $m_3\sim m_2$.

\section{Active neutrino masses}
\label{sec:LHmasses}
At energy scales below the mass of the lightest RHN, the phenomenology of the model can be properly described by the following effective Lagrangian: 
\begin{align}
\label{eq:lowscalelag}
 \mathcal{L}_\mathrm{eff} \simeq \frac{1}{2}\sum_{a,b} \sum_{i,j}\kappa^{(ab)}_{ij} \left(\overline{L_i} \widetilde{\Phi}_a\right)\left(\widetilde{\Phi}_b^{ T}L^c_j\right)+\mathrm{h.c.}\;,
\end{align}
where the Wilson coefficients at the scale $\mu=M_1$ can be calculated in the usual manner by integrating-out the heavy RHNs:
\begin{align}
\label{eq:kappa}
\kappa^{(ab)}\Big\vert_{\mu=M_1}& \simeq \mathbb{Y}^{(a)} \mathbb{M}^{-1} {\mathbb{Y}^{(b)}}^T \;,
\end{align}
with $\mathbb{Y}^{(a)}$ and $\mathbb{M}$ given in Eq.~(\ref{eq:eff_parameters_approx}) (for calculating $\kappa^{(ab)}$, we neglect the contribution from integrating-out $N_3$, which as mentioned above is subdominant).
For rank-1 Yukawa couplings one finds
\begin{align}
	\label{eq:massscales}
	\kappa^{(ab)}\Big\vert_{\mu=M_1} &\simeq \frac{\vec u_L^{(a)} \otimes \vec u_L^{{(b)} T}}{\vec u_L^{(a)} \cdot \vec u_L^{(b)}} \sum_{i,j =1,2} \mathbb{P}^{(ab)}_{ij} \left( \mathbb{M}^{-1} \right)_{ij}\;,
\end{align}
with $\mathbb{P}^{(ab)} \equiv \mathbb{Y}^{(a)^T} \mathbb{Y}^{(b)}$.
Using the notation of Eq.~(\ref{eq:rank1yuk}) we explicitly find:
	\begin{align}
		\label{eq:kappa_at_M1}
		\kappa^{(ab)}\Big\vert_{\mu=M_1}\simeq  \frac{y_a y_b}{M_0}\, \frac{y_1^2 y_2^2}{y^2_a y^2_b}\,\frac{ \vec u_L^{(a)} \otimes \vec u_L^{{(b)} T}}{ \cos\theta_a \cos\theta_b} \times
		\begin{cases}
			1/( 1-\cos^4\theta_L) & a = b, \\
			-\cos^2\theta_L/( 1-\cos^4\theta_L) & a \neq b\;.
		\end{cases}
	\end{align}
Furthermore, to determine the low-energy neutrino parameters, we include quantum contributions to  $\kappa^{(ab)}$. As in the rest of this paper, and due to the large separation between $M_1$ and the energy scale of neutrino oscillation experiments, the dominant quantum contributions to $\kappa^{(ab)}$ can be encoded in the RGE \cite{Babu:1993qv,Chankowski:1993tx,Antusch:2001vn,Grimus:2004yh}:
\begin{align}
\frac{d\kappa^{(ab)}}{d\, \mathrm{log}\, \mu} =\frac{1}{ 16 \pi^2 } \beta_{\kappa^{(ab)}}\;,
\label{eq:RGE-kappa}
\end{align}
with dominant terms of the $\beta$ function at one loop given by:
\begin{align}
\label{eq:betakappaGrimusFlavorless}
	\beta_{\kappa^{(ab)}}&=
	- 3 g^2 \kappa^{(ab)}
	+ \sum_{c,d=1}^{2} 4 \lambda_{acbd} \kappa^{(cd)} 
	+ \sum_{c=1}^{2} \Big[ 3 \Tr\left[Y_u^{(a)}Y_u^{(c)\dagger}\right] \kappa^{(cb)} 
	+ 3 \Tr\left[Y_u^{(b)}Y_u^{(c)\dagger}\right]  \kappa^{(ac)}
	\Big]\;,
\end{align}
where we have neglected all gauge and Yukawa couplings except for the top-quark coupling for the analytical treatment.
The Wilson coefficient of the Weinberg operator at the scale $m_H$ is:
\begin{align}
	\kappa^{(ab)}\Big|_{\mu=m_H}=\kappa^{(ab)}\Big|_{\mu=M_1}+\frac{1}{16\pi^2}\beta_{\kappa^{(ab)}}\log\left(\frac{m_H}{M_1}\right)\;.
\end{align}
Finally, after the electroweak symmetry breaking through the expectation value of the neutral components of the Higgs fields, $\langle \Phi^0_a\rangle=v_a/\sqrt{2}$, with $v_1^2+v_2^2=v^2$, a $3 \times 3$ neutrino mass matrix is generated:
\begin{align}
	\mathcal{M}_\nu \simeq - \frac{1}{2}\sum_{a,b} \kappa^{(ab)}\Big|_{\mu=m_H} v_a v_b\;,
	\label{eq:massmatrix}
\end{align}
where we have neglected the effects of the running between the scale $m_H$ and the scale of the experiment, since the most relevant RGE effects occur between $M_1$ and $m_H$. The impact of quantum effects is twofold. First, each  Wilson coefficient $\kappa^{(ab)}$ receives quantum corrections proportional to itself, which changes the values of the different entries $\kappa^{(ab)}_{ij}$ by ${\cal O}(1)$ factors. Second, and more importantly, the different Wilson coefficients $\kappa^{(ab)}$ can mix through the running, due to ``Higgs changing interactions" in the Weinberg operators, induced by Higgs quartic couplings. This effect is characteristic of the model with an extended Higgs sector, and can significantly affect the low-energy phenomenology \cite{Ibarra:2011gn,Grimus:1999wm,Ibarra:2014pfa,Jurciukonis:2019jkr}.

In order to better differentiate the impact on the phenomenology of the quantum effects above and below the RHN decoupling scale, two scenarios are analyzed. We first discuss a scenario where the operator mixing between the scales $M_1$ and $m_H$ is negligible, and then a scenario where the operator mixing is sizable.

\subsection{Operator mixing between $\kappa^{(ab)}$ negligible}

Following our analysis of Section \ref{sec:RGEs},  we assume that the RHN Yukawa coupling and mass matrices are rank-1 at the cut-off scale $\Lambda$. Then, from Eq.~(\ref{eq:kappa_at_M1}), and neglecting the effects of the running below the scale $\mu=M_1$, one obtains:
\begin{align}
	\mathcal{M}_\nu \simeq \sum_{a,b} \textfrak{m}_{ab} \, \vec u_L^{(a)} \otimes \vec u_L^{(b)T},
	\label{eq:mass_matrix}
\end{align}
where
\begin{align}
	\label{eq:mass_scales}
	\textfrak{m}_{ab} =  \frac{(y_a v_a)( y_b v_b)}{2 M_0}\frac{y^2_1 y^2_2 }{y^2_a y^2_b}\frac{1}{ \cos\theta_a \cos \theta_b} \times
	\begin{cases}
		 1/( 1-\cos^4\theta_L) & a = b, \\
		- \cos^2\theta_L/( 1-\cos^4\theta_L) & a \neq b.
	\end{cases}
\end{align}
The active neutrino mass eigenvalues can be calculated using the tensor invariants:
\begin{align}
	I_1 &= \mathrm{Tr}\big[\mathcal{M}_\nu\big] =m_1+m_2+m_3\;, \nonumber\\
	I_2 &= \frac{1}{2}\big(\mathrm{Tr}\big[\mathcal{M}_\nu]^2- \mathrm{Tr}\big[\mathcal{M}_\nu^2\big]\big)=m_1m_2+m_1m_3+m_2m_3\;,\nonumber\\
	I_3 &= \mathrm{det}\big[\mathcal{M}_\nu\big] = m_1 m_2 m_3\;.
\end{align} 
From Eqs.~(\ref{eq:mass_matrix}) and (\ref{eq:mass_scales}) one obtains:
\begin{align}
	\label{eq:massmatrixinvariants}
	I_1 &=\textfrak{m}_{11} +\textfrak{m}_{22} + (\textfrak{m}_{12} +\textfrak{m}_{21})\cos\theta_L,\\
	I_2 &= \big( \textfrak{m}_{11} \textfrak{m}_{22} - \textfrak{m}_{12} \textfrak{m}_{21} \big) \sin^2\theta_L,\nonumber\\
	I_3 &= 0.\nonumber
\end{align} 
Therefore, $m_1=0$ in the approximation that only two RHNs contribute to the mass matrix.\footnote{Strictly, $m_1$ is non-vanishing although much smaller than the atmospheric and solar neutrino mass scales, since $m_1$ receives contributions at tree level of the order of $v^2/M_3\sim 10^{-5}$ eV. Further, it receives quantum contributions  proportional to $m_3$, although at two-loops~\cite{Babu:1988ig,Choudhury:1994vr,Davidson:2006tg}.}
The other two eigenvalues read, under the assumption $m_3\gg m_2$:
	\begin{align}
		\label{eq:masses_QE_neglected}
		m_3&\simeq I_1 =  
		m_0\,\frac{y_1 y_2}{( 1-\cos^4\theta_L)}\Big[
		\frac{ y_2}{y_1}\frac{\cos^2\beta}{\cos^2\theta_1 } +
		\frac{y_1}{y_2}\frac{ \sin^2\beta }{ \cos^2\theta_2 }
		-\frac{2 \sin\beta\cos\beta \cos^3\theta_L}{ \cos\theta_1 \cos\theta_2} 
		\Big]\;,\nonumber \\ 
		m_2&\simeq \frac{I_2}{I_1}=m_0  y_1 y_2 \sin^2\theta_L \Big[
		\frac{y_2}{y_1}\frac{\cos^2\theta_2}{\sin^2\beta} +
		\frac{y_1}{y_2}\frac{ \cos^2\theta_1}{ \cos^2\beta }
		-\frac{2  \cos\theta_1 \cos\theta_2 \cos^3\theta_L}{\sin\beta\cos\beta} 
		\Big]^{-1}\;,
	\end{align}	
where we have used $v_1 = v \cos \beta$ and $v_2 = v \sin \beta$, and we have defined the overall mass scale
\begin{align}
	m_0=\frac{1}{2}\frac{v^2}{M_0}\;,
\end{align}
with $M_0$ defined in Eq.~(\ref{eq:M0}).
Numerically,
\begin{align}
m_0 \simeq 0.05\,{\rm eV}\left(\frac{M_3}{M_{\rm P}}\right)^{-1}\left(\frac{y_1}{0.7}\right)^{-2}\left(\frac{y_2}{0.7}\right)^{-2}\;,
\label{eq:m0_approx}
\end{align}
which is in the right ballpark if $M_3$ is around the Planck scale, and the largest Yukawa eigenvalues are ${\cal O}(1)$.

It is evident in Eq.~(\ref{eq:masses_QE_neglected}) that a necessary condition to generate a non-vanishing $m_2$ is to have a misalignment between $\vec u_L^{(1)}$ and $\vec u_L^{(2)}$ ({\it i.e.} $\sin\theta_L\neq 0$). Further, the overall scales of $m_2$ and $m_3$ are determined by the same parameter $m_0$. Therefore, for generic values of the misalignment angles, for $y_1\sim y_2$ and for $\tan\beta\sim 1$ one expects a mild hierarchy between $m_2$ and $m_3$. Concretely, the mass hierarchy is estimated to be
\begin{align}
	\label{eq:hierarchySS1}
	\left\vert\frac{m_3}{m_2}\right\vert \sim
	\frac{\left(  \frac{y_2\cos\theta_2\sin\beta}{y_1 \cos\theta_1\cos\beta} +  \frac{y_1 \cos\theta_1\cos\beta}{y_2\cos\theta_2\sin\beta} - 2 \cos^3\theta_L \right)^2}{\sin^2\theta_L \left(1-\cos^4\theta_L \right)},
\end{align} 
which is $\sim 1-10$ under the assumptions listed above.
This result is independent of the alignment of the neutrino Yukawa couplings $Y^{(1,2)}$ with the charged lepton Yukawa eigenbases.
Predictions for the leptonic mixing matrix are hence not possible unless one imposes restrictions on the flavor structure of the Yukawa couplings, {\it e.g.} from flavor symmetries. Generic misalignment angles at high energies will lead at low energies to an ``anarchic" structure for the leptonic mixing matrix, with angles which are neither small nor maximal, in qualitative agreement with observations.

\begin{figure}[t!]
	\begin{center}
		\includegraphics[width=0.6\textwidth]{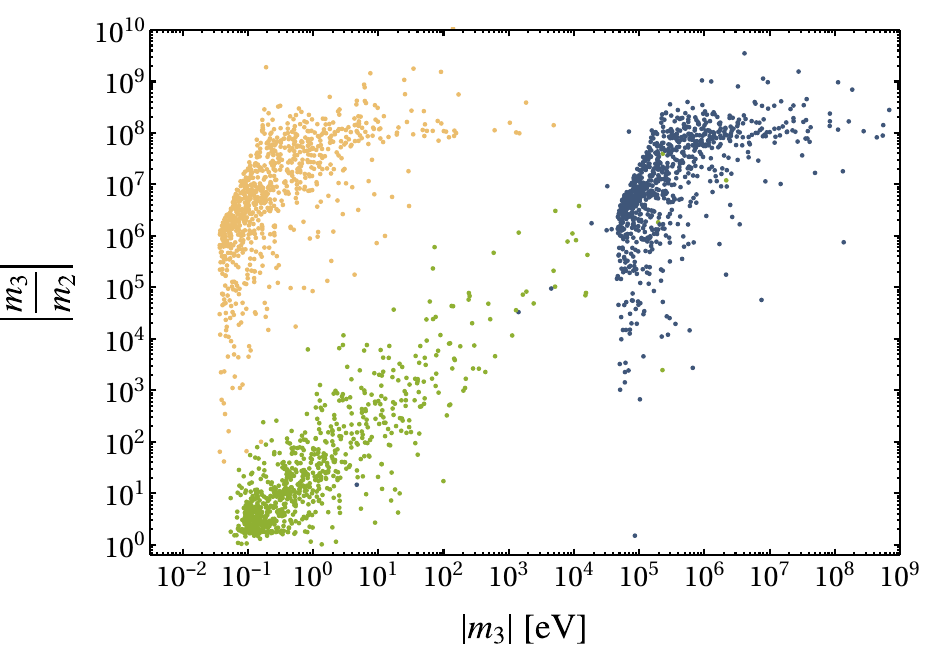}
	\end{center}
	\caption{Scan plot showing the mass hierarchy between the two heavier active neutrinos $\vert m_3/m_2 \vert$  vs. $|m_3|$ in the Planck-scale lepton number breaking scenario with two Higgs doublets, in the case where RGE-induced mixing between the $\kappa^{(ab)}$ is negligible.
	We take $\Lambda=M_{\rm P}$,  $M_3 = M_{\rm P}/\sqrt{8 \pi}$, $y_2=1$, random angles between 0 and $2 \pi$, and $y_1 =1$, $\tan \beta =1$ (green points), $y_1 = 1$, $\tan \beta = 0.01$ (orange points); $y_1 = 0.01$, $\tan \beta = 1$ (blue points).}
	\label{fig:2hdm_no_op_mixing}
\end{figure}

These expectations are confirmed by our numerical analysis. We consider different realizations of our scenario at the cut-off $\Lambda=M_{\rm P}$, assuming  a rank-1 RHN mass matrix with $M_3=M_{\rm P}/\sqrt{8\pi}$ and rank-1 Yukawa matrices with eigenvalues $y_2=1$, and $y_1=1$ or 0.01. The Yukawa eigenvectors $u_{R,L}^{(a)}$ are chosen randomly.
We then solve numerically the two-loop RGEs for the RHN parameters above the scale $M_1$, and the one-loop RGEs for the Wilson coefficients below the scale $M_1$, neglecting the terms in Eq.~(\ref{eq:betakappaGrimusFlavorless}) that mix the different $\kappa^{(ab)}$.
Finally, we calculate the neutrino mass matrix assuming $\tan\beta=1$ or 0.01. The resulting values for  $m_3$  and $|m_3/m_2|$ are shown in the scan plot in Fig.~\ref{fig:2hdm_no_op_mixing}, for the cases {\it i}) $y_1=1$ and $\tan\beta=1$ (green points), {\it ii}) $y_1=1$ and $\tan\beta=0.01$  (orange points), and {\it iii}) $y_1=0.01$ and $\tan\beta=1$ (blue points). When the Yukawa eigenvalues are $y_2\sim y_1 \sim 1$ and $\tan\beta\sim 1$ the predicted neutrino parameters are in the ballpark of the experimental values. However, when $y_1/y_2$ and/or $\tan\beta$ are very different from 1, the predicted neutrino mass hierarchy is generically too large. It is remarkable that this simple scenario can already reproduce the observations for reasonable parameters. Further, and as we will see in the next subsection, the allowed parameter space widens when including the operator mixing induced by quantum effects below the scale $M_1$.

\subsection{Operator mixing between $\kappa^{(ab)}$ non-negligible}

We consider in what follows the phenomenologically interesting case of the 2HDM in the decoupling limit, where the lighter CP-even scalar resembles the Standard Model Higgs, while the other scalars are very heavy. 
In this case, $\tan\beta = v_2/v_1  \simeq 0$, so that ${\cal M}_\nu=\kappa^{(11)} v_1^2/2$, with $v_1=v$. If quantum effects between $M_1$ and $m_H$ were negligible, the decoupling limit would lead to a very large neutrino mass hierarchy, {\it cf.} Eq.~(\ref{eq:hierarchySS1}). However, operator mixing below the scale $M_1$ can significantly modify this conclusion. 

The relevant Wilson coefficient $\kappa^{(11)}$ at the scale $m_H$ is calculated from
$\kappa^{11}|_{m_H}=\kappa^{11}|_{M_1}+\delta\kappa^{11}$ where
\begin{align}
	\delta\kappa^{11}\simeq-\frac{1}{16\pi^2}\beta_{\kappa^{11}}\Big|_{\mu=M_1}\log\left(\frac{M_1}{m_H}\right)\;.
\end{align}
Using the explicit form of the $\beta$ function, this correction can be schematically written as~\cite{Ibarra:2011gn}:
\begin{align}
	\delta \kappa^{11}\simeq
	B_{1a}\kappa^{a1}+\kappa^{1a}B_{1a}^T+ b \kappa^{22}\;,
	\label{correction}
\end{align}
which makes clear the operator mixing through the RGE running.	Here $B_{1a}$ denote $3\times 3$ matrices whereas $b$ is a number. Explicitly,
\begin{align}
	b=-\frac{2 \lambda_5}{16\pi^2} \log\left(\frac{M_1}{m_H}\right)\;,
\end{align}
which depends linearly on the coefficient of the potential
term $\frac{1}{2} \lambda_5 (\Phi_1^\dagger \Phi_2)(\Phi_1^\dagger \Phi_2)$ while only logarithmically on the ratio between the scale of the RHN and the overall scale of the extra scalars $H^0$, $H^\pm$, $A^0$.\footnote{In the notation of Eq.~(\ref{eq:V_2HDM}), $\lambda_5 = 2 \lambda_{1212} = 2 \lambda_{2121}^*$.}
 Due to the large separation of scales between $M_1$ and $m_H$, the large logarithm  $\log(M_1/m_H)$ partially compensates the loop factor, resulting in $b\sim{\cal O}(0.1)$ for $\lambda_5\sim{\cal O}(1)$. Concretely,
\begin{align}
b\simeq - 0.3\,\lambda_5 \,\log\left[\left(\frac{M_1}{10^{14}\,{\rm GeV}}\right)\left(\frac{m_H}{10\,{\rm TeV}}\right)^{-1}\right]\;,
\end{align} 
where we have taken $m_H\gg v$ to implement the decoupling limit.

Expressing $\mathcal{M}_\nu$ in terms of mass parameters $\textfrak{m}_{ab}$ as in Eq.~(\ref{eq:mass_matrix}), we find for the case $\tan\beta=0$:
\begin{equation}
	\label{eq:massParamsL5}
	\begin{split}
		\textfrak{m}_{11} &=  -\frac{m_0}{(1-\cos^4\theta_L)}\frac{y_2^2}{\cos^2\theta_1}, \\
		\textfrak{m}_{22} &=   -\frac{m_0 }{(1-\cos^4\theta_L)}\frac{b\, y_1^2}{\cos^2\theta_2}, \\
		\textfrak{m}_{12} &= \textfrak{m}_{21} = 0.
	\end{split}
\end{equation}
Using the invariants from Eq.~(\ref{eq:massmatrixinvariants}), we obtain for the largest active neutrino mass:
\begin{align}
m_3\simeq  -m_0 \frac{y_1 y_2}{(1-\cos^4\theta_L)}\Big[\frac{y_2}{y_1\cos^2\theta_1}+\frac{b\, y_1}{y_2\cos^2\theta_2} \Big]\;,
\end{align}
and the mass hierarchy
\begin{align}
	\label{eq:hierlambda5}
	\left| \frac{m_3}{m_2} \right| &\simeq \frac{1}{\sin^2\theta_L} \Big[\frac{y_2 \cos\theta_2}{\sqrt{b}\,y_1\cos\theta_1}+\frac{\sqrt{b}\, y_1 \cos\theta_1}{y_2\cos\theta_2} \Big]^2\;.
\end{align}

\begin{figure}[t!]
	\begin{center}
		\includegraphics[width=0.6\textwidth]{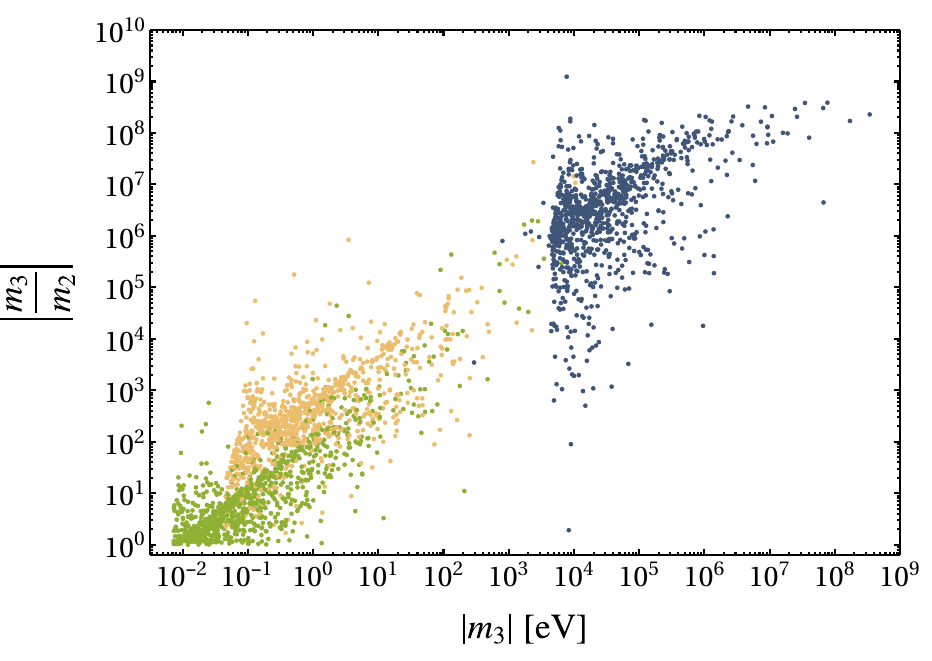}
	\end{center}
	\caption{Same as  Fig.~\ref{fig:2hdm_no_op_mixing}, but for a scenario with $\lambda_5=1$, such that the operator mixing due to running below $\mu=M_1$ is non-negligible.}
\label{fig:2hdm_with_op_mixing}
\end{figure}

Therefore, for $y_1\sim y_2$, generic angles $\theta_1,\theta_2, \theta_L$ and $\lambda_5 \sim \mathcal{O}(1) $ (so that $b={\cal O}(1)$), a mild neutrino mass hierarchy is generically expected.  
The effect of $\lambda_5$ in the running is illustrated in Fig.~\ref{fig:2hdm_with_op_mixing}, which considers the same scenarios as in Fig.~\ref{fig:2hdm_no_op_mixing}, but including the running between $M_1$ and $m_H$ setting $\lambda_5=1$. Clearly, for plausible values of $\lambda_5$ the operator mixing has a significant impact on the phenomenology and widens the allowed parameter space of the model.  

\section{Planck-scale lepton number breaking scotogenic scenario}
\label{sec:scotogenic}
We consider now a variant of the previous scenario where the SM symmetry group is extended with a discrete $Z_2$ symmetry, assumed to be exact in the electroweak vacuum. All SM particles are even under the $Z_2$ symmetry. Further, the SM particle content is extended with fermion singlets, $N_i$, and scalar doublets with identical gauge quantum numbers as the SM Higgs boson, $\eta_a$, all odd under the $Z_2$ symmetry. 

With this set-up, all lepton number violating interactions involving only SM particles vanish at tree level. In particular, the Weinberg operator arises at the one-loop level. Further, the lightest particle of the $Z_2$-odd sector constitutes a dark matter candidate. This is the so-called ``scotogenic model'' \cite{Ma:2006km}. This model, however, presents the same drawbacks as the type-I seesaw model in regards of predicting the correct neutrino mass and neutrino mass hierarchy. It was argued in \cite{Hehn:2012kz,Escribano:2020iqq} that the extension of the scotogenic model by an additional $Z_2$-odd scalar doublet leads in general to a mild neutrino mass hierarchy. In this section, we will investigate whether the breaking of the lepton number at the Planck scale in a variant of the scotogenic model, extended by an additional $Z_2$-odd scalar doublet, can simultaneously lead to the correct neutrino mass and mass hierarchy. 

The part of the Lagrangian containing the $Z_2$-odd fermions and scalars reads:
\begin{align}
 \mathcal{L}_N = \frac{1}{2} \overline{N_i} i \slashed{\partial} N_i - \frac{1}{2} M_{k} \overline{N^c_k}N_k
 -Y^{(a)}_{\alpha i}\overline{L_\alpha} N_i \widetilde\eta_{a} + \mathrm{h.c.}, \label{eq:lagrangianeta}
\end{align}
with $a = 1,2$. Here we have chosen without loss of generality to work in the basis for the singlet fermions where the mass matrix is diagonal with eigenvalues $M_k$. The scalar potential can be split into three separate parts
\begin{align}
	\label{eq:scalarInert}
	V_\mathrm{pot}(\Phi, \eta_1, \eta_2) &= V_\Phi(\Phi) + V_\eta(\eta_1, \eta_2)+ V_\mathrm{int}(\Phi, \eta_1, \eta_2),
\end{align}
with $a=1,2$. Here, $V_\Phi(\Phi) = \mu^2  \Phi^\dagger \Phi + \frac{\lambda}{2} (\Phi^\dagger \Phi)^2$ is the potential for the $Z_2$-even scalar doublet (the SM Higgs doublet),
 $V_\eta$ has the form of Eq.~(\ref{eq:V_2HDM}), replacing $\Phi_a$ by $\eta_a$, and
\begin{align}
V_\mathrm{int}(\Phi, \eta_1, \eta_2) &= \frac{1}{2} \lambda_3^{(ab)} \left( \Phi^\dagger \Phi \right) \left(\eta_a^\dagger \eta_b\right) + \frac{1}{2} \lambda_4^{(ab)} \left(\Phi^\dagger \eta_a\right)\left(\eta_b^\dagger \Phi\right) \\ \nonumber
&+ \frac{1}{2} \lambda_5^{(ab)} \left(\Phi^\dagger \eta_a\right)\left(\Phi^\dagger \eta_b\right) + \mathrm{h.c.}\,,
\end{align}
is the interaction potential between the Standard Model Higgs doublet and the inert doublets.
The masses of the neutral components of the inert doublets will be denoted by $m_{\eta_a}$.

\begin{figure}[t!]
	\begin{center}
		\includegraphics[width=0.45\textwidth]{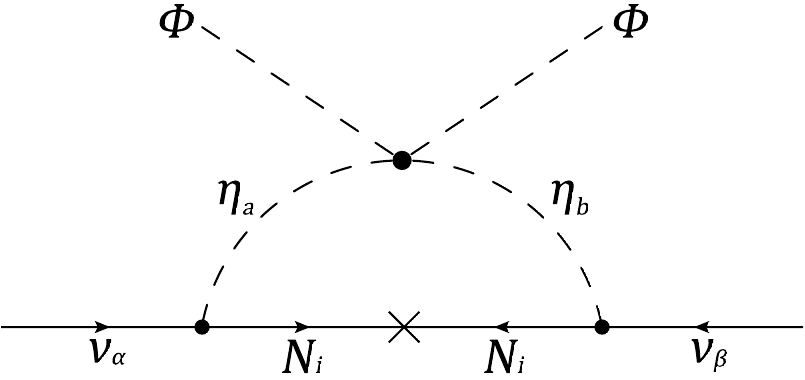}
	\end{center}
	\caption{One-loop diagram generating neutrino masses in the scotogenic scenario with two inert doublets. }
	\label{fig:scoto1L}
\end{figure}

As in the rest of this paper, we assume the RHN mass matrix to be approximately rank-1 at the cut-off scale of the theory, for which we take $\Lambda=M_{\rm P}$. We set for simplicity $M_3\sim M_{\rm P}$ and $M_1,M_2=0$. Two-loop quantum effects induced by the inert doublets generate non-zero values for $M_1$ and $M_2$, given by Eq.~(\ref{eq:RHmass}), with the appropriate substitutions. Integrating-out the heavy particles, a single Weinberg operator arises, shown in Fig.~\ref{fig:scoto1L}, corresponding to the effective Lagrangian
\begin{align}
	\label{eq:lowscale_scotogenic}
	\mathcal{L}_\mathrm{eff} \simeq \frac{1}{2}\sum_{\alpha,\beta}\kappa_{\alpha\beta} \left(\overline{L_\alpha} \widetilde{\Phi}\right)\left(\widetilde{\Phi}^{ T}L^c_\beta\right)+\mathrm{h.c.}\;,
\end{align}
with
\begin{align}
	\kappa_{\alpha\beta} \simeq \sum_{a,b,k}\frac{Y_{\alpha k}^{(a)} Y_{\beta k}^{(b)} \lambda_5^{(ab)} }{16 \pi^2} \frac{M_k}{m_{\eta_b}^2 - M_k^2} \left\{\frac{m_{\eta_b}^2}{m_{\eta_a}^2 - m_{\eta_b}^2} \, \textrm{log}\left(\frac{m_{\eta_a}^2}{m_{\eta_b}^2}\right) - \frac{M_k^2}{m_{\eta_a}^2 - M_k^2} \, \textrm{log}\left(\frac{m_{\eta_a}^2}{M_k^2}\right)\right\}.
\end{align}
Using that in our scenario $M_k\gg m_{\eta_1}$, $m_{\eta_2}$, one can simplify\footnote{This result differs by the one reported in \cite{Ma:2006km} and \cite{Hehn:2012kz} by factors of 2, as pointed out in \cite{Escribano:2020iqq} and \cite{Merle:2015ica}. Note that \cite{Merle:2015ica} defines $v=\langle \Phi^0\rangle$, while \cite{Escribano:2020iqq} takes $v=\langle \Phi^0\rangle/\sqrt{2}$, as we do, hence our formula coincides with the one in \cite{Escribano:2020iqq}.}:
\begin{align}
	\left(\mathcal{M}_\nu\right)_{\alpha\beta} \simeq \sum_{a,b,k}\frac{\lambda_5^{(ab)}}{32 \pi^2} \,  Y_{\alpha k}^{(a)} Y_{\beta k}^{(b)} \frac{v^2}{M_k}\,\textrm{log}\left(\frac{m_\eta^2}{M_0^2}\right)\;,
\end{align}
where in the logarithm  we have approximated  both scalar masses by $m_\eta$, and both fermion masses by $M_0$ (see Eq.~(\ref{eq:M0})).

It is now straightforward to calculate approximate expressions for the largest active neutrino mass, using the rank-1 assumption for the Yukawa and RHN mass matrices as previously
\begin{align}
 \label{eq:scotoLHNmasses}
 m_3 &\simeq {\rm Tr}({\cal M}_\nu)\simeq  -\frac{1}{16\pi^2}\frac{m_0 y_1 y_2}{(1-\cos^4\theta_L)} \left( \frac{y_2\lambda_5^{(11)}}{y_1 \cos^2\theta_1} + \frac{y_1\lambda_5^{(22)}}{y_2 \cos^2\theta_2} - \frac{2\cos^3\theta_L \lambda_5^{(12)}}{\cos\theta_1\cos\theta_2}  \right)\log\left(\frac{m_\eta^2}{M^2_0}\right) \,,
 \end{align}
and for the neutrino mass hierarchy
\begin{align}
 \label{eq:scotoLHNhierarchy}
 \left|\frac{m_3}{m_2}\right| &\simeq \frac{[{\rm Tr}({\cal M}_\nu)]^2}{\frac{1}{2}\vert{\rm Tr}({\cal M}_\nu)^2 - {\rm Tr}({\cal M}_\nu^2)\vert}\simeq
 \frac{\Big[ \frac{y_2}{y_1}\frac{\cos\theta_2}{\cos\theta_1}\lambda_5^{(11)} + \frac{y_1}{y_2}\frac{\cos\theta_1}{\cos\theta_2}\lambda_5^{(22)} - 2\cos^3\theta_L \lambda_5^{(12)}  \Big]^2}{\Big[\lambda_5^{(11)}\lambda_5^{(22)} - \big(\lambda_5^{(12)}\big)^2  \cos^4\theta_L\Big] \sin^2 \theta_L} \,.
\end{align}
The overall neutrino mass scale is suppressed with respect to $m_0$, given in Eq.~(\ref{eq:m0_approx}), as well as the loop factor, but also enhanced by the large logarithm $\log( m_\eta^2/M_0^2)$. Therefore, when the relevant couplings $y_1$, $y_2$ and $\lambda_5^{(ab)}$ are all ${\cal O}(1)$, and for generic misalignment angles, one again finds an overall neutrino mass scale in the ballpark of the experimental values. Similarly to our conclusions in Section \ref{sec:LHmasses}, it is necessary to have $\sin\theta_L\neq 0$ ({\it i.e.} a misalignment between the vectors $\vec u_L^{(1)}$ and $\vec u_L^{(2)}$) in order to generate a non-vanishing $m_2$. 
These results of the neutrino mass eigenvalues are again independent of the misalignment between neutrino and charged lepton Yukawa couplings. Taking this misalignment to also be generic, one expects a leptonic mixing matrix with mixing angles which are neither small nor maximal, in qualitative agreement with observations.

\begin{figure}[t!]
 \begin{center}
  \includegraphics[width=0.5\textwidth]{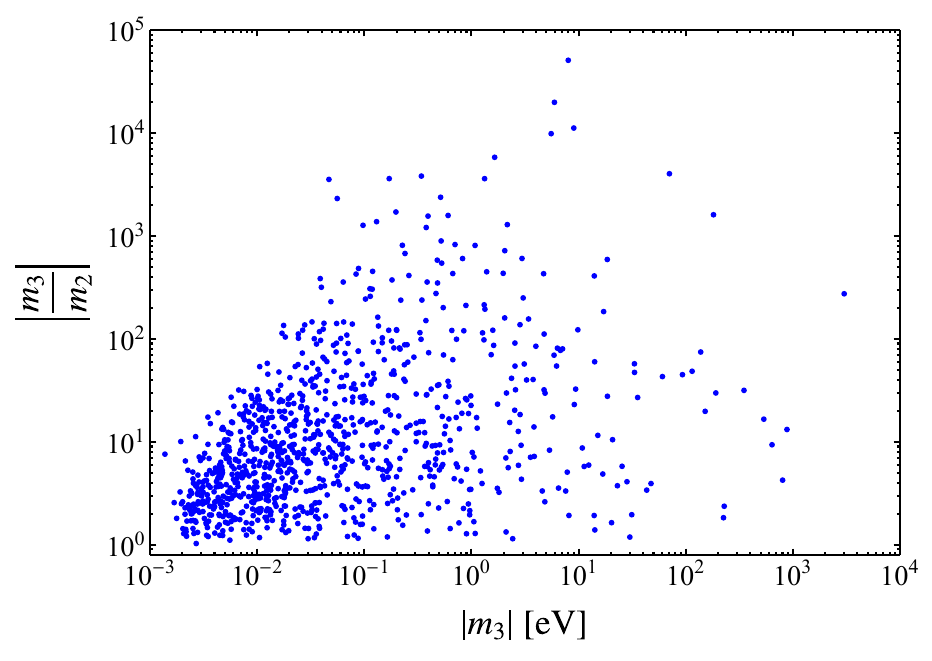}
 \end{center}
 \caption{Scan plot showing the mass hierarchy between the two heavier active neutrinos $\vert m_3/m_2 \vert$  vs. $|m_3|$ in the scotogenic Planck-scale lepton number breaking scenario with two inert doublets,  assuming $\Lambda=M_{\rm P}$,  $M_3 = M_{\rm P}/\sqrt{8 \pi}$, and inert doublet masses  $m_{\eta_1} = 100 \, \mathrm{TeV}$ and $m_{\eta_2} = 2 m_{\eta_1}$. The Yukawa couplings have  eigenvalues $y_1 = y_2 = 1$  and random misalignment angles, and the quartic couplings are $\lambda_5^{(11)}=\lambda_5^{(22)}=0.1$ and all other quartic couplings are taken to be zero for simplicity.}
 \label{fig:hierarchyscoto}
\end{figure}

Our conclusions are illustrated in Fig.~\ref{fig:hierarchyscoto}, which shows the largest neutrino mass and mass hierarchy for a random scan of the flavor directions for the scotogenic scenario, assuming $\Lambda=M_{\rm P}$,  $M_3 = M_{\rm P}/\sqrt{8 \pi}$, $m_{\eta_1} = 100 \, \mathrm{TeV}$ and $m_{\eta_2} = 2 m_{\eta_1}$, as well as $y_1=y_2=1$, and  $\lambda_5^{(11)} = \lambda_5^{(22)} = 0.1$, $\lambda_5^{(12)} = \lambda_5^{(21)} = 0$. As anticipated, most points lie in the region with $|m_3|=0.001-0.1$ eV, and $|m_3/m_2|<100$.

\section{Conclusions}
\label{sec:conclusion}

We have considered an extension of the type-I seesaw model by extra scalar doublets, assuming that the total lepton number is broken by a right-handed neutrino mass close to the Planck scale. In this setup, lighter right-handed neutrino masses receive contributions at two loops, proportional to the mass of the heaviest. In this work we have focused on the scenario where the lighter masses are dominated by these quantum contributions. This scenario, that we denominate Planck-scale lepton number breaking seesaw scenario, has a larger predictive power compared to the general seesaw framework, since the whole neutrino mass spectrum is determined by only two mass scales, the Planck scale and the electroweak symmetry breaking scale, which are known.

We have shown that under fairly general conditions, this scenario leads to  active neutrino masses in the ballpark of the experimental values. At very high energies, two-loop quantum effects induced by the two Higgs doublets, generate comparable masses for the lighter right-handed neutrinos. Integrating-out the right-handed neutrinos leads to small neutrino masses through the seesaw mechanism. One of the active neutrinos is predicted to be $m_3\sim (16\pi^2)^2 v^2/M_{\rm P}\sim 0.1$~eV. Further, the mild hierarchy between the two lighter right-handed neutrino masses generically leads to mild hierarchies between the two largest eigenvalues of the Wilson coefficients of the Weinberg operators. This already suggests the generation of solar and atmospheric mass scales with a mild hierarchy. Further quantum effects due to the operator mixing among the Weinberg operators assist in the generation of a mild neutrino mass hierarchy, which then becomes a fairly generic expectation of the model.
Moreover, one expects angles in the leptonic mixing matrix which are neither small nor maximal, although the precise values cannot be predicted in our phenomenological approach, which does not impose any restriction on the flavor structures at high energies.
Let us also stress that this scenario does not require a light exotic Higgs sector and the same conclusions apply in the decoupling limit, where lepton flavor and CP-violating processes have suppressed rates. 

We have finally considered a ``scotogenic" variant of this scenario with three fermion singlets and two scalar doublets carrying a $Z_2$ charge. The same conclusions apply for the neutrino phenomenology. Further, the ``inert" doublets in this case make for a viable dark matter candidate.

\section*{Acknowledgments}

This work has been supported by the Collaborative Research Center SFB1258 and by the Deutsche Forschungsgemeinschaft (DFG, German Research Foundation) under Germany's Excellence Strategy - EXC-2094 - 390783311.

%

\bibliographystyle{JHEP}
\bibliography{refs}

\end{document}